\documentstyle[revtex]{aps}
\begin{document}
\begin{center}

\Large\bf
Centrality  of collisions and total disintegration of
nuclei.

\vskip 5mm
M.K. Suleimanov$^{1 \ast}$, O.B. Abdinov$^{1}$, A.I. Anoshin$^{2}$,
J.Bogdanowicz$ ^{3}$ and A.A.Kuznetsov$^{4}$.

\vskip 5mm

{\small
(1) {\it
Physics Institute, Azerbaijan Academy of
Sciences, Baku, Azerbaijan Republic and
the Laboratory of High Energies, JINR
}
\\
(2) {\it
Nuclear Physics Institute, Moscow State University, Moscow, Russia
}
\\
(3) {\it
Soltan Institute for Nuclear Studies, Warzaw, Poland
and the Laboratory of High Energies, JINR
}
\\
(4){\it
The Laboratory of High Energies, JINR
}
\\
$\ast$ {\it
E-mail: mais@sunhe.jinr.ru
}}
\end{center}

\vspace {1.cm}

\section {\bf Introduction.}
{ \large

The search for signals connected with superdense states of
nuclear matter is one of the basic trends of research in
experiments on relativistic nuclear physics [1]. The best conditions of
research of such states are the studies of events with a maximum number
of nucleons - participants in the interaction  or events connected
with central collisions of nuclei.
To select  such events,
the following criteria are usually used: events with maximum
multiplicities of secondary particles or events with a minimum
flow of energy of secondary particles emitted at  a zero angle (see Refs.
[2], the  results presented in these papers show much importance
of studying the central collisions for a full understanding of the
processes of interactions of relativistic nuclei). Theoretically,
both of these conditions correspond to the value of impact parameter
$b\rightarrow 0 $.  The centrality  of nucleus - nucleus collisions is
really the best condition for arising superdense states of nuclear
matter. However, this condition is not  sufficient as there are
processes with a high degree of centrality of collisions, but they do
not result in arising superdense states of nuclear matter [4]
(besides,the $ b $-dependence of the production cross section of
superdense states of nuclear matter can be of a resonance character).
In this cases,it is necessary to introduce an additional condition of
selection of such a type of events to observe a signal from superdense
states of nuclear matter. We think that such conditions can be from  the
research of processes with total disintegration of nuclei  in
interactions of relativistic nuclei [4]. For this, it is necessary to
relate processes with total disintegration of nuclei  to cases with a minimum flow of energy of particles emitted at a zero  angle as  the condition  of a minimum flow of energy of particles emitted at a zero  angle  is now in use as a basic trigger for central collision selection .

The research of the processes with total disintegration of nuclei
started long ago in experiments with emulsion nuclei [5].  Interest in
them was primarily connected with that anomalously  high
densities of nuclear matter could be realized in these processes and the
effects, related to collective properties of nuclear matter,
could be observed.  However, contrary to the expectations, one could not
receive in the experiment an unequivocal answer to the
question on the realization of these states.
 To our mind, the reasons of this are the following: 1) the absence
of an adequats insight into the  kind of signals of superdense states
of nuclear matter; 2) in the above experiments there was no opportunity
to take into account the cases, in which large momenta were transferred to
fragments in interactions; the energy characteristics of secondary particles
were not practically determined, and the statistical material, as a rule,
did not exceed some hundred events.

Taking into account all this and also the importance of
the problem,  the processes with total disintegration of nuclei were
studied in our experiment [6-8] according to a new experimental
statement. It included the following:

 (a) The bubble chamber technique was used that allowed
the energy and the charge sings  of all secondary
particles to be determine.

 (b) The development of new selection criteria  of events with
total disintegration of nuclei.  For this purpose, the idea is used that
the processes with total disintegration of nuclei
correspond to qualitatively  new states of  nuclear matter and
the transition to these states occurs in nuclear interactions
when the number of  protons emitted from nuclei, $ Q $, reaches a
critical value of $ Q ^ * $, at which  the regime change happens in the
behaviour of the characteristics of secondary particles in
$ Q $-dependences .  Hence one can use the following condition as a
selection criterion for events with total disintegration of nuclei :

\begin {equation}
Q\ge Q ^ *
\end {equation}

 This method of selection of reactions with total disintegration of
nuclei is experimentally realized by studying the behaviour of
different characteristics of secondary particles in nucleus - nucleus
interactions depending on $ Q $. So,  the $ Q $-dependences of the
following characteristics are considered [6-8] :  probabilities to
observe events, the average characteristics and inclusive spectra of
secondary particles, and also one-particle correlation functions for
$\pi $ - mesons and protons.  The results, obtained in these papers ,
have confirmed the assumption of the  existence a certain boundary value
of $ Q^* $ ( which excess leads to the processes with total
disintegration of nuclei ).  The experimental data on the dependence of
the average multiplicity of relativistic  charged particles of the sum
of charges of projectail fragments for $ {} ^ {28} Si _ {14} + Em$ (at
14.6 A GeV) and $ Si + Em $ (at 3.7 A GeV) reactions are presented in
paper [9].  The regime change was observed  in these dependences at the
transition from the region of large values of the sum of charges of
projectile fragments the region of small values  - to the region of
central collisions.   The values of the sum of charges of projectile
fragments corresponding to the points of regime change in these
dependences were used to select the central collisions.  We believe that
at our energies this result can mean the existence of the transition of
nuclear matter from nucleon states to its non-nucleon and mixed states.
At  RICH or LHC energies , a similar result can mean the detection of
"critical" points of phase transition to nuclear matter, and  it can be
used  to develop adequate representations of  the kind of a  signal from
superdense states of nuclear matter.}

\section { Methods of the experiment.}

{ \large

 The experimental data, obtained in an exposure of the
2-m propane bubble chamber to relativistic nuclei at a momentum of 4.2 A
GeV/c, were used in the analysis. The total statistics of events
are:  8130 events - $ pC $  , 6945 - $ dC $ , 11248 - $ {} ^ 4HeC $
and 20407 - $ {} ^ {12} CC $  interactions. Methodical
details are described in [9]. It should be noted, that protons
in this experiment are reliably identified by ionization and  path only
over a momentum interval of  0.15-0.50 GeV/c.  Protons with a momentum
  of $ p < 0.15 $ GeV/c have a path shorter than 2 mm and most of them
are not seen in the photograph.  The weights, determining the
probability that the given particle is a proton or a $ \pi ^ + $ -
meson, are assigned to all positive particles having a momentum  higher
than 0.5 GeV/c .  The characteristics of $ \pi ^ - $ - mesons were used
to determine the  weights.  The minimum momentum for the detection of
$ \pi ^ - $ -mesons was 0.07 GeV/c.  The fraction of electrons and
negative strange particles did not exceed 5 $ \% $ and 1 $ \% $,
respectively.  To determine the number of protons (as well as in ref.
[6-8]), the variable $ Q $ was used. The value of  $ Q $ for each event
was determined as

\begin {equation}
Q = N _ {+} - N _ {\pi ^ {-}}.
\end {equation}

\noindent Here $ N _ {+}  $ and $ N _ {\pi ^ {-}} $ are the numbers  of
positive particles and $ \pi ^ {-} $ - mesons,respectively
(assuming that $ N _ {\pi ^ {+}} = N _ {\pi ^ {-}} $).
The experimental losses of particles and
errors in identifying secondary particles and fragments affect the accuracy
of determining the values of $Q$.
A bad accuracy in determining $ Q $ can result in the
appearance of "false" $ Q ^ * $ and extension of the regions of
regime changes.
For this reason  we cannot
determine precisely the number of regions of regime change  and the
values of $ Q ^ * $ corresponding to them.
To decrease the
influence of this factor, we consider not groups of events with
definite values of $ Q $ and groups of
events with  $ Q $ larger than a certain value, i.e. the integral
spectrum.
 Under such  consideration, the influence of accidents of
all kinds decreases.
Therefore, the experimental
material was divided into groups of events with the following values of
$ Q $:

\begin {equation}
Q\geq 1; 2; 3;.. Q ^ {max}.
\end {equation}

\noindent For example, we took  11 for $ Q ^ {max} $ for
$ {} ^ {12}CC $
interactions and $ Q ^ {max} = 7 $ for the other types of interactions .}

\section {\bf Results.}
{ \large

As already noted, the basic aim  of the present work is to
relate events with total disintegration of nuclei to cases with a minimum flow of energy of particles emitted at a zero  angle.
To achieve this
aim in the experiment, it was supposed
that if these events correlated, with increasing $ Q $ the
average values of

\begin {equation}
K = {\sum _ {i = 1} ^ {n} {p ^ 2 _ i} \over \sum _ {i = j} ^ {N} {p ^ 2 _ j}}
\end {equation}

\noindent ( Here $ p ^ 2 _ i $ is the
momentum squared of the  charged particles with an emission
angle $\theta \le 5^0$  in the laboratory system of coordinates
and $ n $ is the  number of these particles; $ p ^ 2 _ j $  is the
momentum squared of all charged particles and $ N $ is their number
in the event) must decrease sharply and  reach
a minimum value for the events with a minimum flow of energy at emission
angles close to $0^0$.
In this case, if the value of $ <b> $ decreases with increasing $ Q $,
this means an approach to the condition of central collision
(to define the values of $< b >$
we uced the calculation data by the quark-gluon string model ( QGSM)
[10]).

 Fig.1. shows the Q-dependences of $ < K > $ .   It is seen
that the values of  $ <K > $ decreasing with increase $ Q $ :  for $ {}
^{12} CC $ interactions in the interval $ Q\ge 6 $, for $ {} ^ 4HeC $ -
$ Q\ge 4$, and for  $ dC $, $ pC $ - interactions in the interval $ Q\ge
3 $.  Thus, one can conclude that in the interval of large $ Q $, i.e.
in area of total disintegration of nuclei,  $ <K> $ decreases with
increasing $ Q $ and reaches its  minimum at a maximum value of $ Q $.
 From here it follows that the events with total disintegration of nuclei
correspond to the cases with a minimum flow of energy of charged particles at an emission angle of $\theta \le 5 ^ 0 $.

 One can also see (fig.2) that with increasing $ Q $,  $ <b> $
decreases (calculated data on QGSM) and reaches its minimum at a
maximum value of $ Q $.  This means that the processes with total
disintegration of nuclei, selected with the help of the condition
$ Q\ge Q ^ * $ in the framework of QGSM, correspond to events with the
highest centrality of collisions. One can also  see that  the values of
$ <b> $ increase with growing $ A $.  Thus, the events with total
disintegration of nuclei, selected with the help of the condition
$ Q\ge Q ^ * $, correspond to the cases with the largest centrality or the
cases with a minimum flow of energy at an angle of $0^0$.  In so doing,
the determined values of $ < K ^* > $ correspond to the values of
$ Q ^ * $.

 From the data in  fig.1., one can see  that there is a
strong $ A $ - dependence for the distributions $ <K> = f (Q) $. A
similar result has been obtained in  [7]. As  was
noticed there, one-partial correlations weaken with  increasing $ A $,
and  they become minimum for $ {} ^ {12} CC $ interactions.
As it follows from  [7] and the present data, the character of
dependence of correlation functions on $ Q $ also changes simultaneously
with weakering correlations in the region $ Q \geq Q ^ * $ ( total disintegration of nuclei).
 In the [8], it has also been found that there are two regions
in the behaviours of back slopes of invariant inclusive spectra in
$ Q $-dependences .
We have interpeted this fact as an indication of
a probable growth of the density (or " temperature ") of nuclear matter
in the area of total disintegration of nuclei.
 Thus, the results, obtained in this paper, come  the
 conclusion that  the condition $ <K> < < K ^ * > $ (corresponding to
 the condition $ Q \geq Q ^ * $) can be used (as an additional one) in
the experiments, studying the reactions with a minimum flow of energy of
secondary particles at a zero angle, for  experimental  detection of a
signal from superdense  states of  nuclear matter.  For this, the
opportunity of an application of event-by-event  analysis  is discussed
[12]. We have no opportunity for direct application of the
event-by-event  analysis, as in our experiment the multiplicity of
secondary particles  is much smaller than it is necessary for
the event-by-event analysis. But we believe that
 the method of event analysis used by us in the present work
and the obtained results will bee important in the
event-by-event  analysis (to receive a signal from quark-
gluon plasma) in collisions of heavy nuclei at high energies. The
event-by-event analysis must give  more full information on
the dynamics of nuclear collisions than the inclusive
analysis. To search for fluctuations with the  event-by-event
analysis, it is necessary to exclude changes of collision geometry.
Event selection with fixed values of $ b $ is
assumed to be the best way for it .   The values of $ b $ can
   be estimated through the values of a flow of energy in
 fragmentation regions, in particular through a flow of energy at an
 angle  $ 0 ^ 0 $.    Thus, information on the energy and volume
 dependence of the obtained experimental results is needed for
an unambiguous interpretation of the obtained data.   The
   energy dependence can be taken into account by  comparison of
 the data obtained at different energies of colliding nuclei.  The
 volume dependence of results can be taken into account by
 comparison of the data obtained for collisions of nuclei with
 different masses.  It is expected that in the event-by-event
 analysis the cases with quark gluon plasma will differ from those
 without plasma in the  point of regime change in the corresponding
 dependences. The first results of the NA49 Collaboration [13] in the
 event-by-event  analysis for Pb + Pb collisions at SPS energies
 were reported. }

\vspace{1.0cm}

\section {Acknowledgement}

{ \large

    The authors consider it their pleasant debt to thank
the staff of the two-meter propane bubble chamber for the
given experimental material and also Acad. A.M. Baldin for
his constant attention to our work and Prof. H.M. Zinovyev for
useful discussions and notes.

The research is supported by Grant INTAS-96-0678.
}

Fig.1. $ Q $ - dependence of
the values of  $ < K > $ for $ {} ^ {12} CC $, $ {} ^ 4HeC $, $ dC $ and
$ pC $ interactions.

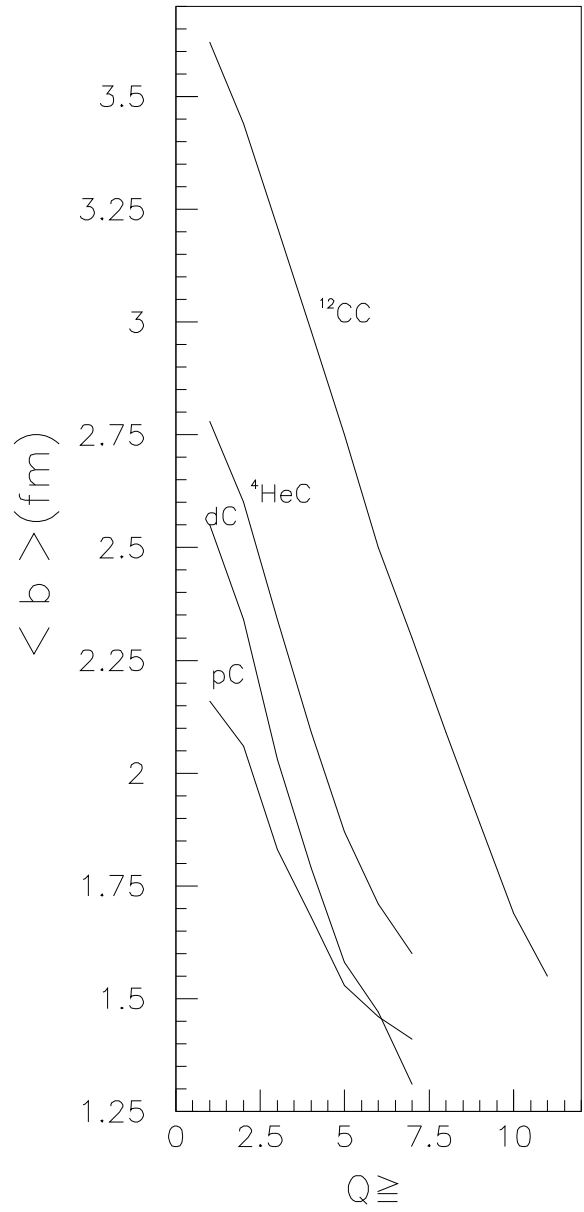
Fig.2. $ Q $ -
dependence of the values of  $ < b > $ for $ {} ^ {12} CC $, $ {} ^ 4HeC
$, $ dC $ and $ pC $ interactions.

\newpage
\begin{center}
{\bf References}
\end{center}
\begin {itemize}

\item [1]. J. W. Harris and STAR Collaboration, in: Proceedings X
International Conference on Ultra-Relativistic Nucleus-Nucleus
Collisions, Borlange, Sweden, June 20-24, 1993; Nucl. Phys., A566 (1994)
277;

J.Schukraft and ALICE Collaboration, in: Proceedings X
International Conference on Ultra-Relativistic Nucleus-Nucleus
Collisions, Borlange, Sweden, June 20-24, 1993; Nucl. Phys., A566 (1994)
311.

\item [2.] .
 J. Barrette et al. (E814  Collaboration)  Phys. Rev. C50(1994) 3047;

     G. Wang et al. (E900  Collaboration)  Phys. Rev. C53(1996) 1811;

       Y. Akiba et. al. (E802  Collaboration)  Phys. Rev. C56(1997) 1544;

     B. Hong  et al. (FOPI  Collaboration) Phys. Rev. C57 (1998) 244;

      Miskowiec et. al (KaoS  Collaboration)  Phys. Rev. Lett. 72 (1994)
3650;

      W.C. Hsi et al. (ALA DIN Collaboration) Phys. Rev. Lett. 73 (1994)
3367;

     M.M. Aggarwal et al. (WA80,WA93,WA98 Collaboration) Phys. Rev. C56 (1997) 1160;

      T. Alber et al. (NA35 and NA49 Collaborations) Nucl. Phys. A590
(1995) 453;

      J. Bachler et.al  (NA35 Collaborations) Z. Phys. C58 (1993) 541;

      H. Appelshauser et al. (NA49 Collaborations) Eur.Phys.J.C2 (1998)
661;

Chkhaidze et al. Physics Letters B, volume 411, N 1,2, p.26-32,1997.;

L. Ahle, Y. Akiba, et al.
Phys. Rev. C, volume 55, Number 5, p.2604-2614.1997.

\item [3.] Waged Kh., Uzhinskii V.V. JINR Communications N E2-94-126,
Dubna, 1994; K.K. Gudima, V.V. Toneev, Preprint JINR P2-10431, Dubna
1977 ; Sov. Journal Nucl. Ph. v.27, 11, 1978.

\item [4.]
  Akhrorov et al.Preprint JINR P1-9963, Dubna, 1976.; K.D. Tolstov, R.A.
  Khoshmukhametov. Preprint JINR, P1-6897, Dubna, 1973;  B.Jakobsson,
  J.Otterlund, K.  Kristiansson. Preprint LUIP-CR-74-14, Lund, 1974;
  AA-B-G-D-D-E-K-K-M-P-SP-S-T-T-UB-U-Collaboration. Sov.Journal Nucl.
Ph.F, v. 55, 4, 1992, p.1010-1020; Bagdanov V. G. et al. Sov.Journal
  Nucl.Ph. v .38, 1983 , p. 1493; Yu. F.Gagarin et al. Sov.Journal "
  News of USSR Academy of Sciences " Phys. Series, v. 38,
N 5,1974, p. 989- 992 ; N.Angelov et al., Sov. Journal Nucl.Ph. v.28, 3
  (9), 1978.; Bondarenko A.I. et al. Sov. Journal Nucl. Ph. v.60, 11,
  1997.; A.  Dabrowska et al.  Phys. Rev. D47 (1993) 1751.

\item [5]. V.A. Belyakov et al. Preprint JINR, P - 331, Dubna, 1959.

\item [6]. Abdinov O.B. et al. JINR  Rapid Communications, N 1[75],1996;
Abdinov O.B. et al. JINR  Rapid Communications, N 7[81],1997;

\item [7].
  Abdinov O.B. et al. JINR Communications,
E1-97-342, Dubna, 1997; hep-ex/9712025.

\item [8]. Abdinov O.B. et al. JINR Communications, E1-97-178, Dubna,
1997.

\item [9.]
M.I. Tretyakova. EMU-01 Collaboration. Proceeding of the XIth
International Seminar on High Energy Physics Problems. Dubna, JINR,
1994.,p.616-626.

\item [10]N.Akhababian et al.- JINR Preprint 1-12114, Dubna, 1979.;
N.S.Angelov et al.- JINR Preprint 1-12424, Dubna, 1989.

\item [11] N.S. Amelin, L.V. Bravina. Sov. Journal Nucl. Ph. 51, p.211,
1990.; N.S. Amelin et al., Sov. Journal Nucl. Ph. 50, p. 272, 1990.

\item [12.]  Marek Gazdzicki,Andrei Leonidov and Gunther Roland.
University of Frankfurt.   Preprint IKF-HENPG/5-97.;

E.V. Shuryak. State University of New York. Preprint SUNY-NTG-97-17.

\item [13.]  Preceedings H. Feldmeier, J.Knoll, W. Nwrenberg and J.
Wambach  GSI, Darmstadt, 1997 (309-318).

\end {itemize}

\end {document}